%% file: 0_main.tex
\documentclass[manuscript]{acmart}
\AtBeginDocument{%
  }

\usepackage{framed}
\usepackage{enumitem}
\usepackage{url}

\acmISBN{978-1-4503-XXXX-X/2018/06}

\begin{document}

\title[Incentive-Tuning: Understanding and Designing Incentives for Empirical Human-AI Decision-Making Studies]{Incentive-Tuning: Understanding and Designing Incentives for Empirical Human-AI Decision-Making Studies}

\author{Simran Kaur}
\affiliation{%
  \institution{Delft University of Technology}
  \country{Delft, The Netherlands}
}
\email{skaur1@tudelft.nl}

\author{Sara Salimzadeh}
\affiliation{%
  \institution{Delft University of Technology}
  \country{Delft, The Netherlands}
}
\email{s.salimzadeh@tudelft.nl}

\author{Ujwal Gadiraju}
\affiliation{%
  \institution{Delft University of Technology}
  \country{Delft, The Netherlands}
}
\email{u.k.gadiraju@tudelft.nl}


\begin{abstract}
While artificial intelligence (AI) has revolutionised decision-making across various fields, human judgement remains paramount for high-stakes decision-making processes. This has fueled explorations of collaborative decision-making between humans and AI systems, aiming to leverage the strengths of both.
To explore this dynamic, researchers conduct \textit{empirical studies}, investigating how humans use AI assistance for decision-making and how this collaboration impacts results.
A critical aspect of conducting these studies is the \textit{role of participants}, often recruited through crowdsourcing platforms. The validity of these studies hinges on the behaviours of the participants, hence effective incentives that can potentially affect these behaviours are a key part of designing and executing these studies.
Incentive schemes can thus be the bridge between the controlled environment of the study and the complexities of real-world decision-making. By carefully designing incentives that align with the study goals and participant motivations, researchers can unlock the true potential of empirical studies for investigating human-AI decision-making. 
In this work, we aim to address the critical role of incentive design for conducting empirical human-AI decision-making studies, focusing on \textit{understanding}, \textit{designing}, and \textit{documenting} incentive schemes.
Through a thematic review of existing research, we explored the current practices, challenges, and opportunities associated with incentive design for human-AI decision-making empirical studies. We identified recurring patterns, or \textit{themes}, such as what comprises the components of an incentive scheme, how incentive schemes are manipulated by researchers, and the impact they can have on research outcomes.
Leveraging the acquired understanding, we curated {a set of guidelines} to aid researchers in designing effective incentive schemes for their studies, called the Incentive-Tuning {Framework}, outlining how researchers can undertake, reflect on, and document the incentive design process.
By advocating for a standardised yet flexible approach to incentive design and contributing valuable insights along with practical tools, we hope to pave the way for more reliable and generalizable knowledge in the field of human-AI decision-making.
\end{abstract}




\maketitle

\input{1_Introduction}
\input{2_Method}
\input{3_Results}
\input{4_Discussion}
\input{5_Framework}
\input{6_Implications}
\input{7_Conclusions}


\bibliographystyle{ACM-Reference-Format}
\bibliography{references}


\end{document}

%% file: 1_Introduction.tex
\section{Introduction and Background}
\label{sec:intro}

To effectively integrate AI into human decision-making processes, it is important to develop a fundamental understanding of how humans interact with AI systems, how they incorporate AI advice into their decision-making, and how this collaboration impacts outcomes.
To do so, researchers have been conducting empirical studies that investigate the dynamics of human-AI interaction in the context of decision-making, exploring different factors such as algorithmic aversion, trust, reliance, fairness perceptions, explainability, cognitive biases, and more \cite{overcomingAA2018, AccuracyOnTrust19, SeemsSmartActsStupid23, perceptionsofjustice2018, DisparateInteractions19, ConfidenceAndExplanation2020, 48, ThereIsNotEnoughInformation22, DecidingFastandSlow22, ToTrustOrToThink21, ValueSimilarity23}.
These studies often simulate real-world decision-making scenarios, with participants playing the role of human decision-makers and providing valuable insights into how humans interact with AI systems and the effectiveness of AI assistance \cite{DresselFarid2018, overcomingAA2018, HumanAgentTeam2016}. Some studies also task participants with assessing the quality of AI-generated recommendations or explanations, or with providing feedback on AI systems and interfaces in order to identify areas for improvement in AI models \cite{perceptionsofjustice2018, progressivedisclosure19, DecisionMakingUI2019, 40}.

{Crowdsourcing marketplaces have enabled different research communities to conduct large-scale experiments requiring human participation in various capacities \cite{CSGhezzi2017,gadiraju2017crowdsourcing,peer2017beyond}}.
The design of crowdsourced studies involves several key aspects, such as crafting and allocating tasks to the crowd, providing incentives to participate, implementing quality control measures, and aggregating individual contributions into usable data that can yield valuable insights \cite{CrowdsourcingComponents18, CrowdsourcingElements21, Incentives2009, QualityControlCS18}.
Among these, incentive schemes serve as the linchpin for motivating and retaining participants as well as ensuring active engagement and maintaining data quality \cite{Incentives2009, IncentiveSurvey18}.
When participants feel fairly compensated and motivated by incentives, they could be more likely to truly engage with the tasks, and emulate the real-world motivations of the human whose role they are playing \cite{MotivIncent17, Monetary78}.
The design of incentive schemes is in itself a field of interest, with researchers exploring how to appropriately reward crowd workers while ensuring the reliability and validity of study outcomes within various domains \cite{SurveyCS11, IncentiveSurvey18}.
Common approaches include monetary rewards, such as payment per task or hourly rates, as well as non-monetary incentives like gamification elements, badges, or platform recognition for high-quality contributions \cite{IncentiveSurvey18, GamificationinCS16, AchievementSys14}.
Researchers have also focused on optimizing monetary incentive structures to strike a balance between motivating participation and controlling costs \cite{or11, ICML13, budgetfixaamas14}, while also considering ethical implications such as fair compensation and preventing exploitation \cite{EthicsCS16, EthicalCS2018}.
Hence, understanding and implementing effective incentive schemes is imperative for the success of any crowdsourced study, including empirical studies for human-AI decision-making.

While traditional crowdsourcing tasks often comprise of simple microtasks like data annotation or transcription \cite{CrowdsourcingMTurk, CrowdsourcinginCV16}, decision-making tasks involve complex cognitive processes \cite{complexcognition2010}. Participants of human-AI decision-making studies often contribute to high-stakes decision-making processes, analyzing information, making judgments, and assessing AI capabilities, ultimately impacting the very foundation of our understanding of human-AI interaction within decision-making \cite{GroundingPaper23}.
Designing crowdsourcing experiments that reflect the real-world stakes of such tasks can be challenging, since participants are usually recruited from platforms where the primary incentive is monetary \cite{RunningExpsMTurk10}. Thus, the monetary incentive --- and not the perceived stakes of the decision itself --- can become the primary motivator for the decision-makers.
%
How researchers motivate and incentivize participants can influence the inferences of the results of a study \cite{IntrinsticExtrinsic18}. Some prior work has discussed the possibility of participant motivation affecting certain results, acknowledging it as the potential cause behind observations \cite{DresselFarid2018, ExplanationsCanReduceOverreliance23}. 

Despite these nuances, the current research landscape relies heavily on simplistic \cite{ImperfectAI19, ExplainingModels19, CooperativePlay19, 35, 44, CounterfactualExplanations22, CapablebutAmoral22, ThereIsNotEnoughInformation22, Howdoyoufeel23} or ad-hoc \cite{AdhocIS17, DresselFarid2018, overcomingAA2018, ExplainabilityAccountability20, 48, DoHumansTrustAdvice22, ConditionalDelegation22, WhoShouldITrust23} {monetary} incentive schemes{, which do not describe or necessarily capture the specific context, research goals, and desired outcomes}. While few studies discuss the potential impact of incentives for their experiments \cite{BailingJailing19, TakingAdvicefromDisSimilarMachines22, DecidingFastandSlow22, KnowingAboutKnowing23, 81, ExplanationsCanReduceOverreliance23, HumansForegoRewards23}, there are also studies that do not describe incentive schemes at all \cite{PerceptionsofJustice18, HumanInterptblty18, AIModeratedDecisionMaking22}. Incentive schemes thus seem to rarely be strategically tailored to align with the research goals or to address the unique challenges presented by the field of human-AI decision-making, which go beyond simply identifying an "appropriate" pay amount, but also include identifying the desired behaviours to be encouraged, evaluated, and incentivised.  
The variations and inconsistencies in the design and descriptions of incentive schemes within prior work highlight a critical gap --- current practices \textit{lack a standardised approach} to the design as well as the documentation of incentives.
This lack of cohesion has also caused the research community to suffer from a \textit{fragmented understanding} of the role of incentives. This has created a significant knowledge gap, as we currently lack even an understanding of the nature of incentive schemes and the potential ramifications associated with them. This "unknown unknowns" situation makes it challenging to evaluate the implications of existing research findings and impedes the development of standardised practices for the future. {Since monetary incentives have been the primary means of incentivizing participants in crowdsourced human-AI decision-making studies, we scope our work to monetary incentives.}

Thus, there is a pressing need for a more nuanced approach towards understanding and designing incentive schemes. Firstly, it is important to foster a deeper understanding of the current scenario of incentive design for human-AI decision-making studies.
Secondly, we need to determine how to actually design incentive schemes, while addressing the unique challenges that the domain presents. Here, we also need to move towards standardization of the design process.
Thirdly, we need to identify how to best document the incentive schemes, their design process and the associated outcomes, so that we can pave the way for future researchers to build upon existing knowledge.
Consequently, we address the following research questions (RQs) {to tackle these important methodological challenges} in this work:

\begin{framed}
    \textbf{RQ1}: How are {monetary} incentive schemes currently designed for conducting empirical human-AI decision making studies?

    \textbf{RQ2}: How can {monetary} incentive schemes be appropriately designed through a standardised process for empirical human-AI decision-making studies?

    \textbf{RQ3}: How can the design of {monetary} incentive schemes be documented through a standardised process to facilitate future research in human-AI decision-making?
\end{framed}

To address these RQs, we conducted a \textit{semi-structured review} and \textit{qualitative analysis} of human-AI decision-making literature by means of a \textit{thematic literature review} - a \textit{reflexive thematic analysis} \cite{TAClarkeandBraun} of \textit{excerpts} describing incentive schemes or discussing participant motivation within published research articles.

The themes identified through this {comprehensive} analysis form the basis of our contribution towards addressing this objective and to the field of human-AI decision-making. Our critical discussion and reflection on the themes culminated in several insights and directions for future work. 
 
We further leveraged these insights to develop the Incentive-Tuning {Framework}, to guide researchers in tuning incentive schemes for their human-AI decision-making studies. {This framework is a practical contribution that can help in establishing standardised incentive practices and enhancing the reliability of human-AI decision-making research. Our work complements ongoing efforts to standardize crowd work study design~\cite{ramirez2021state,oppenlaender2024state,whiting2019fair}.}
We supplement the {framework} with a reporting \textit{template} as well as a public-access collaborative repository for storing and accessing incentive schemes and the rationales behind incentive design decisions for published research. 
All the supplementary materials and resources associated with this work are captured in a public GitHub repository.\footnote{\url{https://github.com/simrankaur1509/IncentiveTuning}}

%% file: 2_Method.tex
\section{Methodology}
\label{sec:method}

\subsection{Scope of Review \& Inclusion Criteria}
In this work we focus on \textit{empirical}, \textit{human-subject} studies that investigate human-AI collaboration in \textit{decision-making tasks}. These are studies that aim to evaluate, understand, or improve human performance and experience within the decision-making context.

Further, since the research questions are centered around exploring incentive scheme design, we also aimed to only investigate studies that recruit crowd workers to play the role of the human decision-makers. Since crowd workers are mostly incentivised with monetary rewards, our focus for further analysis is also scoped to monetary incentives only.

The inclusion criteria for the literature review was thus defined as follows:
\begin{itemize}[nosep]
    \item The investigated tasks must be centered around decision-making activities in human-AI collaboration. Papers focusing on tasks with different goals, such as debugging, model improvement, co-creation, or gaming, are excluded.
    \item The study must involve human decision-makers specifically in the context of crowdwork. Papers that solely focus on in-house personnel to act as the human decision-makers, such as employees of an organization or students of a university, are excluded.
    \item The studies must have an \textit{evaluative} focus, assessing the effectiveness, usability, or impact of human-AI collaboration, with crowd workers having a \textit{task} to perform. This excludes formative, purely qualitative interview or survey based studies.
\end{itemize}


\subsection{Search Strategy}
We identified that \cite{GroundingPaper23} have shared a list\footnote{\url{https://haidecisionmaking.github.io/}} of research papers that was compiled considering an inclusion criteria that is a superset of the inclusion criteria defined for this work.  
Thus, we included all 81 studies presented within it to be screened again against our specific criteria. This covered the potentially relevant literature until the year 2021.
To ensure comprehensiveness for more recent research, we conducted a search focused on publications between 2021 and November 2023. Our search targeted the proceedings of key venues in the field, including the ACM Conference on Intelligent User Interfaces (ACM IUI), the AAAI/ACM Conference on AI, Ethics, and Society (AIES), the ACM Conference on Human Factors in Computing Systems (CHI), the ACM Conference on Fairness, Accountability, and Transparency (ACM FAccT), the ACM SIGCHI Conference on Computer-Supported Cooperative Work \& Social Computing (CSCW), and the AAAI Conference on Human Computation and Crowdsourcing (HCOMP).
The search keywords were "human-AI collaboration" and "human-AI decision-making".

\begin{figure}
    \centering
    \includegraphics[width=0.55\textwidth]{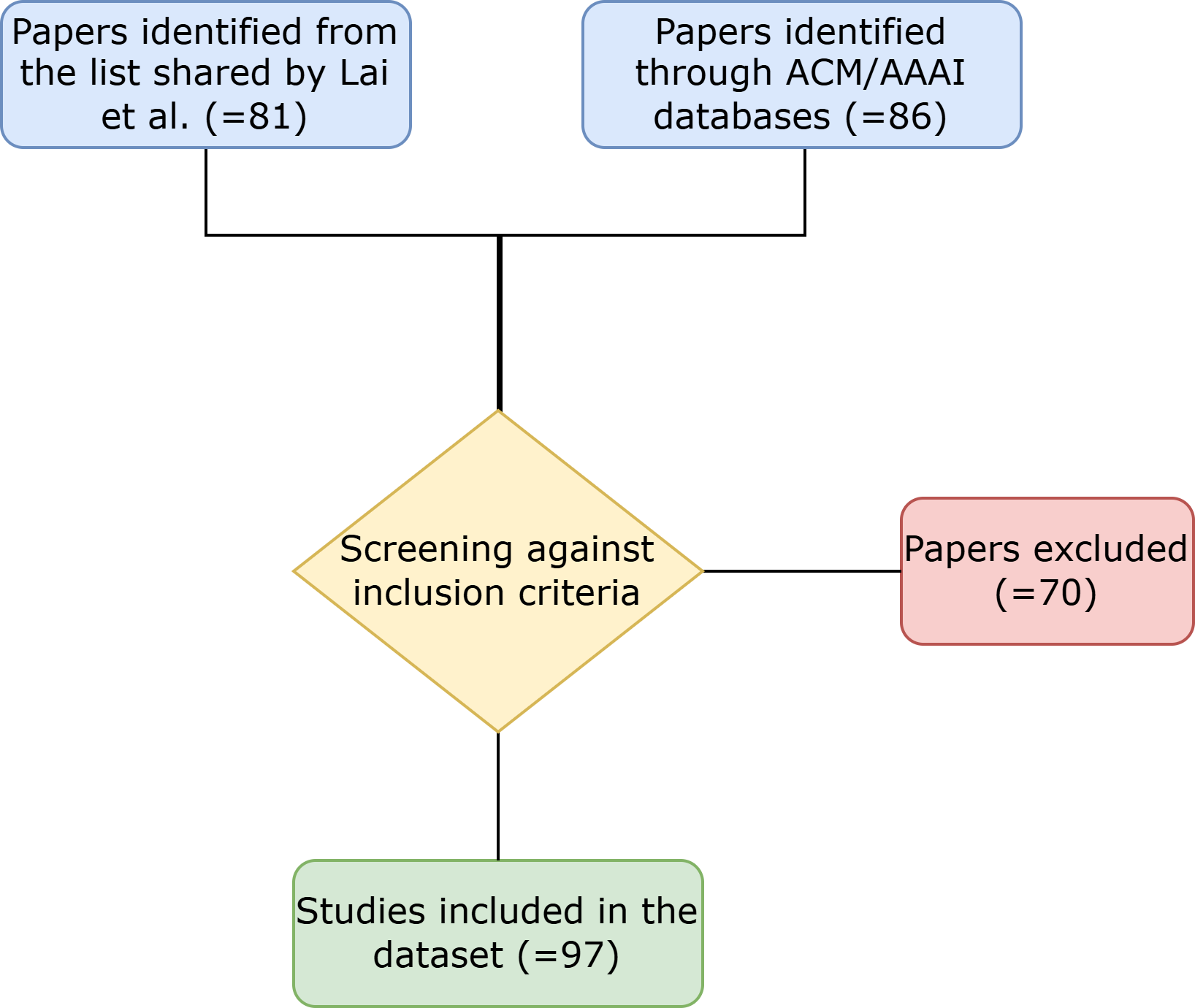}
    \caption{Flowchart illustrating the paper selection process in creating the dataset for thematic analysis}
    \label{fig:literaturesearch}
\end{figure}

An initial screening using titles and abstracts identified 86 potentially relevant papers, resulting in a total of over 160 papers papers to be evaluated further.
Following the initial screening, we meticulously re-evaluated each paper against the defined inclusion criteria. This rigorous assessment resulted in a final selection of 97 papers deemed in-scope for this study. Figure \ref{fig:literaturesearch} illustrates the sequential steps of searching, screening, and including research papers for the thematic analysis. 

\subsection{Dataset Development}\label{Ch2:Dataset}
We first compiled the bibliographic information of the selected papers into a spreadsheet. This captured details such as title, authors, publication year, venue, and an accessible link for each included paper.
Then, for each study, the following information was extracted for each paper: the objective and research goals of the study, the study (and task) domain, the perceived risk or stakes associated with the task, study setup details such as the intended audience and participant information (number of participants, platform, filtering criteria, etc.), the role of participants and the task they had to perform, summary of the employed incentive scheme and task completion time, the excerpts describing the incentive scheme, and any excerpts containing discussions on participant motivation or incentive design.

The \textit{excerpts} describing the incentive design and any discussions surrounding incentives served as the \textit{dataset} for the thematic analysis. 
The rest of the information was deemed to be related to incentive design, and compiled in order to potentially facilitate future analysis.
This dataset is made available for public access \href{https://anonymous.4open.science/r/IncentiveTuning-5D52/Literature%20Review%20Dataset/Literature%20Review%20-%20HAIDM%20studies%20for%20IncentiveTuning.pdf}{here}.

\subsection{Procedure}\label{Ch2:DataAnalysis}
For qualitative analysis of the dataset of excerpts, we conducted a \textit{reflexive thematic analysis}.
Reflexivity is a conscious effort to acknowledge the researcher's role and its potential influence on how data is interpreted. Further, \cite{TAClarkeandBraun} emphasize that reflexivity goes beyond self-reflection - it is a critical examination of both the knowledge produced from the research and how we produce it. 
Thus, in order to practice reflexivity, we explicitly identified our approach to the thematic analysis as \textit{inductive} (allowing themes to directly emerge from within the data), \textit{semantic} (focusing on the explicit content of the data to analyze how incentive design is operationalised and discussed within the excerpts), and \textit{critical} (aiming to unpack the broader meaning and implications of incentive design).
Further, we executed the six phases of reflexive thematic analysis - 'data familiarization', 'data coding', 'initial theme generation', 'theme development and review', 'theme refining, defining and naming', and 'writing up', as described by Braun and Clarke \cite{TAClarkeandBraun}. {The generated codebook can be found} \href{https://anonymous.4open.science/r/IncentiveTuning-5D52/Thematic%20Analysis/Codebook.xlsx}{here}.
Throughout the process, we also followed the "15-point Thematic Analysis Checklist" provided by \cite{TAOG2006} which outlines key criteria for conducting high-quality thematic analysis. This checklist, as applied to the TA conducted for this work, can be found \href{https://anonymous.4open.science/r/IncentiveTuning-5D52/Thematic%20Analysis/Braun%20and%20Clarke%20(2006)%20Checklist%20of%20Criteria%20for%20Good%20Thematic%20Analysis%20Process%20-%20IncentiveTuning.pdf}{here}.

\subsection{Positionality}\label{sec:TAPositionality}
Thematic analysis hinges on a researcher's ability to systematically extract meaning from qualitative data. However, researchers themselves are not blank slates. Their background, experiences, and biases - encapsulating their positionality - can influence how they interpret and analyze the data \cite{TAClarkeandBraun}.
In order to successfully conduct \textit{reflexive} thematic analysis, it is important for the researcher to acknowledge their position. This allows the researcher to be aware of their potential personal perspectives and yet aim to conduct a critical analysis.
Thus, we acknowledge our position by means of a positionality statement given below.

\subsubsection{Positionality statement}
We, the authors who conducted this analysis, are a diverse group of individuals representing different genders, nationalities, and professional backgrounds in academia and industry.
Primarily, our backgrounds lie in various aspects of human-computer interaction (HCI) and artificial intelligence (AI). Through our involvement in human-AI decision-making projects, we have firsthand experience in studying and designing incentive schemes for the kind of studies that we have analysed in this work. Our prior experience has directly exposed us to the complexities of human behaviour within this domain and the challenges of aligning participant motivations within human-AI collaborative studies. Witnessing these dynamics has given us perspective on how well-designed or ill-designed incentives can influence these studies.

In undertaking this thematic analysis of incentive design in human-AI decision-making studies, we acknowledge our positionality and its potential impact. While we believe that possessing a blend of theoretical knowledge and practical experience in this domain has allowed us to approach this research with a comprehensive lens, it may also have led to preconceived notions about how incentive design is, or should be, undertaken within such studies.
{Our research design, specifically the selection of studies included in our analysis, was informed by our existing knowledge and expertise. This approach, while efficient, may have inadvertently excluded studies that did not align with our initial research questions or theoretical framework. Additionally, the power dynamics between researchers and participants in the studies we analysed could have influenced the reporting of incentive schemes, potentially leading to a skewed representation of practices.
Further, a vested interest in human-AI collaboration may have shaped our interpretation of the findings. We may have been more inclined to highlight the positive impacts of well-designed incentives on participant engagement and data quality, while downplaying potential negative consequences, such as ethical concerns or unintended biases.
To mitigate these potential biases, we have rigorously adhered to reflexive thematic analysis methodologies, actively seeking to identify alternative perspectives and challenge our own assumptions. By acknowledging our limitations and actively working to address them, we strive to provide a balanced and insightful analysis of incentive design in human-AI decision-making studies.}

%% file: 3_Results.tex
\section{Thematic Analysis}
\label{sec:thematic_analysis}

The following themes have been identified after performing a rigorous reflexive thematic analysis on the dataset of excerpts that describe incentive schemes in human-AI decision-making literature. We also accompany the descriptions of the themes with specific examples from the literature, as guided by the \cite{TAClarkeandBraun} reflexive TA checklist. 

\subsection{Theme 1: Components of an incentive scheme}
\begin{figure}
    \centering
    \includegraphics[width=0.65\textwidth]{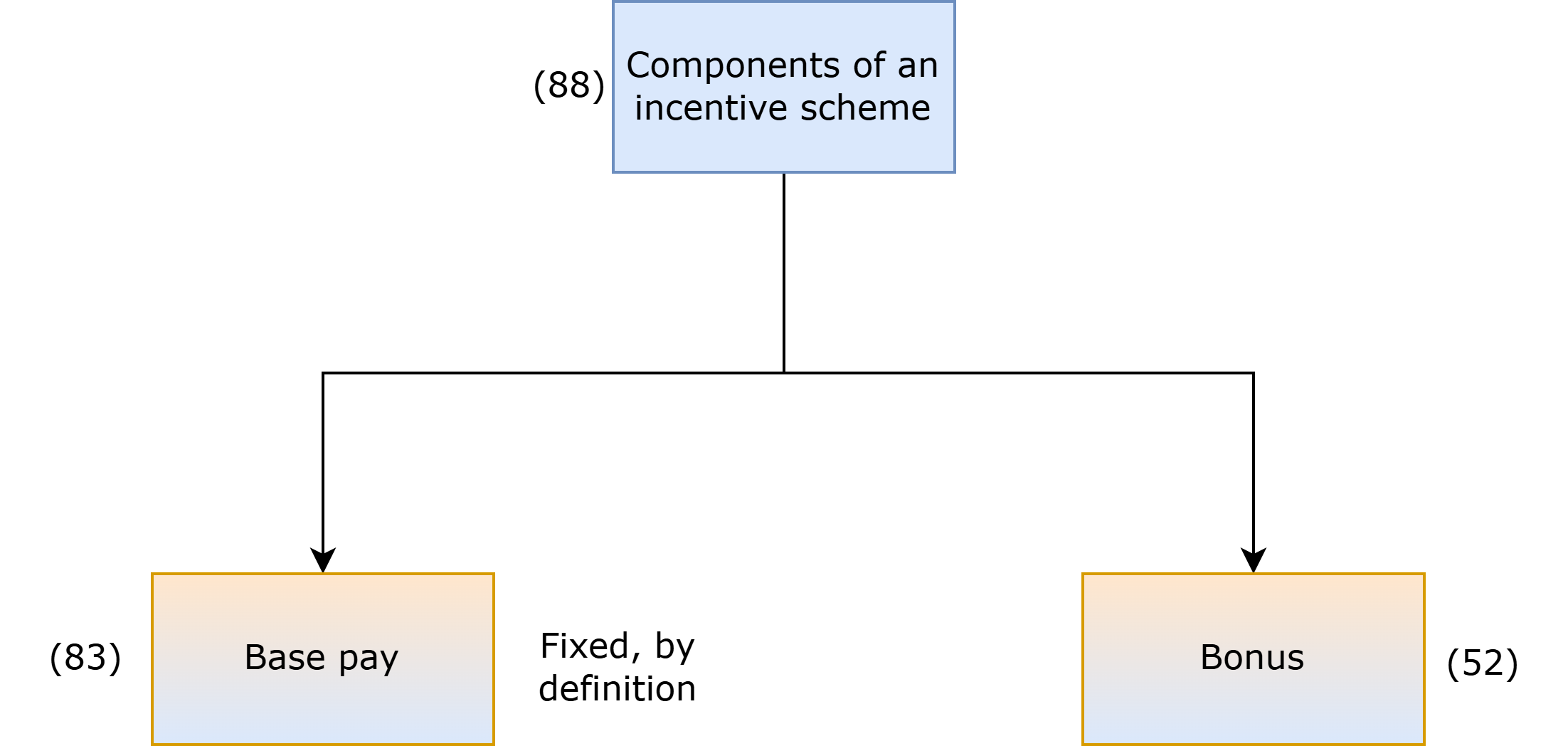}
    \caption{Top-level thematic chart for Theme 1: Components of an incentive scheme}
    \label{fig:Theme1:Components}
\end{figure}

Occurring in 88/97 papers, the components of an incentive scheme is the most common theme. We further identified two sub-themes within it, namely \textit{base pay} and \textit{bonus}.
These are shown in Figure \ref{fig:Theme1:Components}.

47/88 papers mentioned both a base pay and a bonus.
For instance, \cite{2} mentioned, \textit{"Participants were each given \$3 for completing the study (\$1 base and a \$2 bonus to motivate performance)..."}. Similarly, \cite{3} wrote, \textit{"Participants received \$1 for completing the study and they could earn up to an additional \$1 for accurate forecasting performance."}

In 41/88 papers, researchers only mentioned a base pay and not a bonus. For instance, \cite{47} wrote, \textit{"Each participant received a flat payment of \$2.50."}
Meanwhile, 5/88 papers only mentioned a bonus, and not a base pay, like \cite{5}: \textit{"...we offered (relatively) large bonuses to the two annotators who made the most virtual money."}

\subsubsection{Base Pay}\label{Ch2:TA:Results:BasePay}
\begin{figure}
    \centering
    \includegraphics[width=0.8\textwidth]{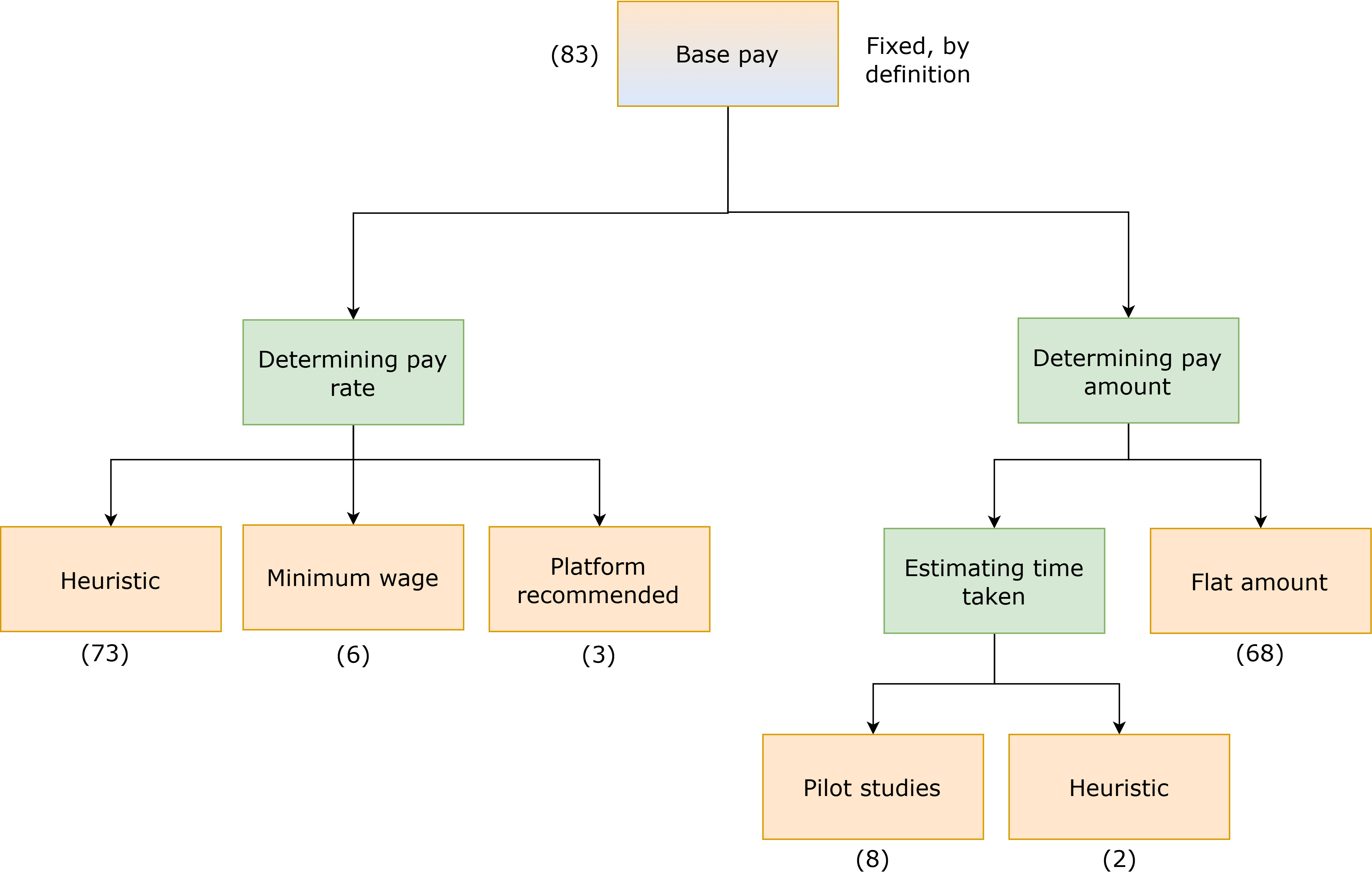}
    \caption{Thematic chart for Theme 1.1: Base Pay and its sub-themes}
    \label{fig:Theme1.1}
\end{figure}

83/97 papers mentioned base pay. Figure \ref{fig:Theme1.1} captures this theme and its sub-themes.

Base pay is identified as the core or basic payment that participants receive for completing the tasks assigned to them. To describe base pay, researchers either directly identify a \textit{pay amount} or they identify a \textit{pay rate}, i.e., a certain amount per hour, or both. 

For instance, \cite{35} wrote, \textit{"Each worker was paid 2 USD."} Meanwhile \cite{82} mentioned, \textit{"Participants were rewarded based on a \$12 hourly rate..."}. \cite{44} described both: \textit{"£3.75 per completion (=£7.09/hour)"}

Pay rates and amounts are most frequently determined \textit{heuristically}, with no explanations given for how the amount was decided. In a few cases, the pay rate is explicitly said to be informed by \textit{minimum wage}, or \textit{platform recommendations}. 
For example, \cite{65} mention, \textit{"To provide fair compensation to our participants, Mturkers were offered an equivalent of United States federal minimum wage..."} while \cite{97} mention, \textit{"All participants were rewarded with [...] deemed to be “good” payment by the platform..."}.

Researchers usually just present a \textit{flat amount}. In a few cases, researchers mention the \textit{estimated time taken} to complete a task along with the pay amount or rate. Estimation is mostly done through \textit{pilot studies}, and in a couple of cases, \textit{heuristically}.
For instance, \cite{79} mentioned: \textit{"From our pilot studies with our research group, we determined that the average time for completing the task was approximately an hour. Therefore, we set a compensation of \$17 for the task."}

\subsubsection{Bonus}
52/97 papers mentioned bonuses. 
A maximum possible bonus \textit{amount}, per task or for the whole study, is decided \textit{heuristically}. For instance, \cite{40} mentioned that, \textit{"Each participant was compensated \$2.50 and an additional \$0.05 bonus for each correctly labeled test review."} Meanwhile \cite{81} mentioned \textit{"All participants were paid \$7.50, including the base value \$5.50 plus a \$2.00 bonus payment..."}.

A \textit{payout scheme} outlines how the final bonus amount which is to be paid to the participants is to be calculated.
These schemes can be classified into different \textit{types}, based on the criteria which is used to determine the payout. These are identified as \textit{performance-based}, \textit{completion-based}, and \textit{luck-based}.

\begin{figure}[!h]
    \centering
    \includegraphics[width=0.75\textwidth]{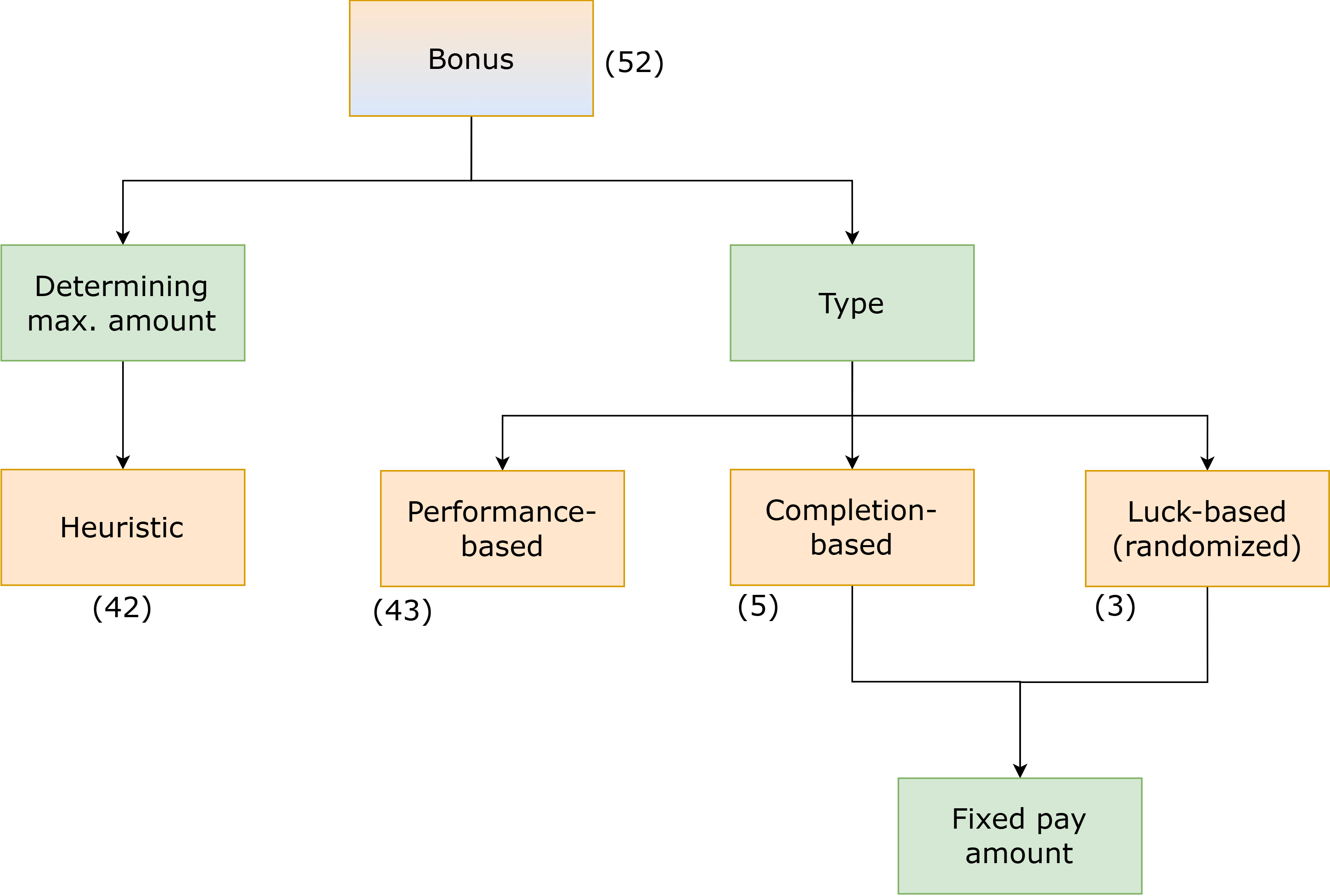}
    \caption{Thematic chart for Theme 1.2: Bonus and top-level sub-themes}
    \label{fig:Theme1.2}
\end{figure}

Figure \ref{fig:Theme1.2} illustrates the high-level theme and sub-themes of bonuses. 

\textit{Performance-based} schemes are the most commonly used. Participants are paid based on their performance, which is evaluated using certain \textit{performance metrics}. A policy for \textit{mapping performance to pay} is usually also presented. For instance, \cite{DresselFarid2018} noted that, \textit{"The participants were paid \$1.00 for completing the task and a \$5.00 bonus if their overall accuracy on the task was greater than 65\%..."}. 

For \textit{completion-based} bonuses, a fixed amount is paid when a participant completes a specific task, such as responding to optional surveys. For instance, \cite{48} mentioned, \textit{"Participants received [...] a fixed bonus of \$0.25 for completing the survey..."}.

For \textit{luck-based} bonuses, there is usually a \textit{randomised} element to determining whether a participant receives a bonus. Randomization is introduced within the design in different ways, such as introducing a specific chance for a participant to earn a bonus, or picking one task at random for each participant to evaluate based on some criteria. For example, \cite{50} mentioned, \textit{"...we randomly selected one prediction task in the sequence to check whether the subject’s final prediction on that task was correct. If so, the subject would receive a \$1 bonus on top of the base payment."}

Further, there are sub-themes that emerge specifically within performance-based bonuses. These are captured in Figure \ref{fig:Theme1.2:Perf-based}, and discussed below:

\begin{figure}[!h]
    \centering
    \includegraphics[width=0.85\textwidth]{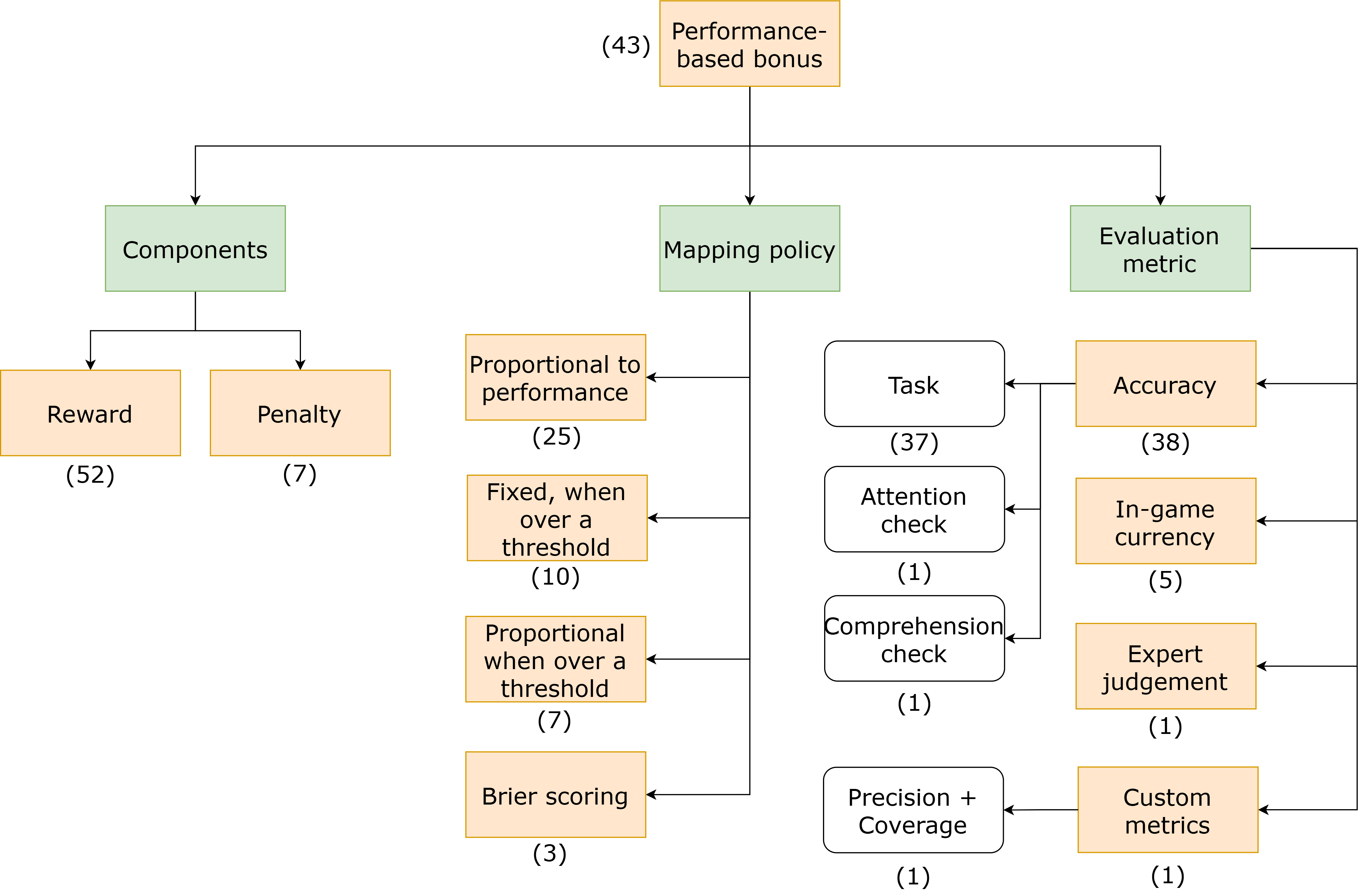}
    \caption{Thematic chart for \textit{performance-based bonus} and its sub-themes}
    \label{fig:Theme1.2:Perf-based}
\end{figure}

Firstly, \textit{performance evaluation metrics} are used to evaluate participant performance, with the most common being \textit{accuracy}. For instance, \cite{41} mentioned, \textit{"...participants received an additional performance-based bonus of £0.5 for each correct answer..."}

Other examples of metrics include in-game currency (gamified score for game-based tasks) \cite{55}, custom defined metrics such as 'precision + coverage' \cite{ConditionalDelegation22}, and expert judgements (experts evaluating participant performance) \cite{6}.

Secondly, there are several ways to \textit{map performance to pay}. These include policies such as fixed payout when performance is over a threshold \cite{DresselFarid2018}, payout proportional to the performance \cite{40}, payout proportional to the performance when performance is over a threshold \cite{52}, and payout calculated through the Brier scoring \cite{Brier} function \cite{19}. For instance, \cite{overcomingAA2018} describe their policy as, \textit{"Participants were paid a \$0.50 bonus if their official forecasts were within five percentiles  of students’ actual percentiles. This bonus decreased by \$0.10 for each additional five percentiles of error [...]. As a result, participants whose forecasts were off by more than 25 percentiles received no bonus."}

Further, a performance-based bonus can also have a \textit{penalty} (negative incentive) alongside a \textit{reward} (positive incentive). For instance, \cite{ConfidenceAndExplanation2020} mentioned, \textit{"...a reward of 5 cents if the final prediction was correct and a loss of 2 cents if otherwise..."}.
The design specifications of such policies are often heuristic, with only a couple of cases being grounded in prior literature \cite{ConfidenceAndExplanation2020, 19}.

{The prevalence of both base pay and bonus in a significant proportion of studies (47/88) suggests that a hybrid approach is commonly employed to incentivize participants. This suggests that researchers tend to gravitate towards a pay structure described by a baseline compensation while also attempting to motivate participants through performance-based rewards.
Further, the high prevalence of heuristic approaches in determining base pay rates and amounts (83/97) indicates a lack of standardised practices in the field. This suggests a need for more systematic and evidence-based approaches to setting compensation levels.}



\subsection{Theme 2: Manipulation of incentives}
\begin{figure}[!h]
    \centering
    \includegraphics[width=0.85\textwidth]{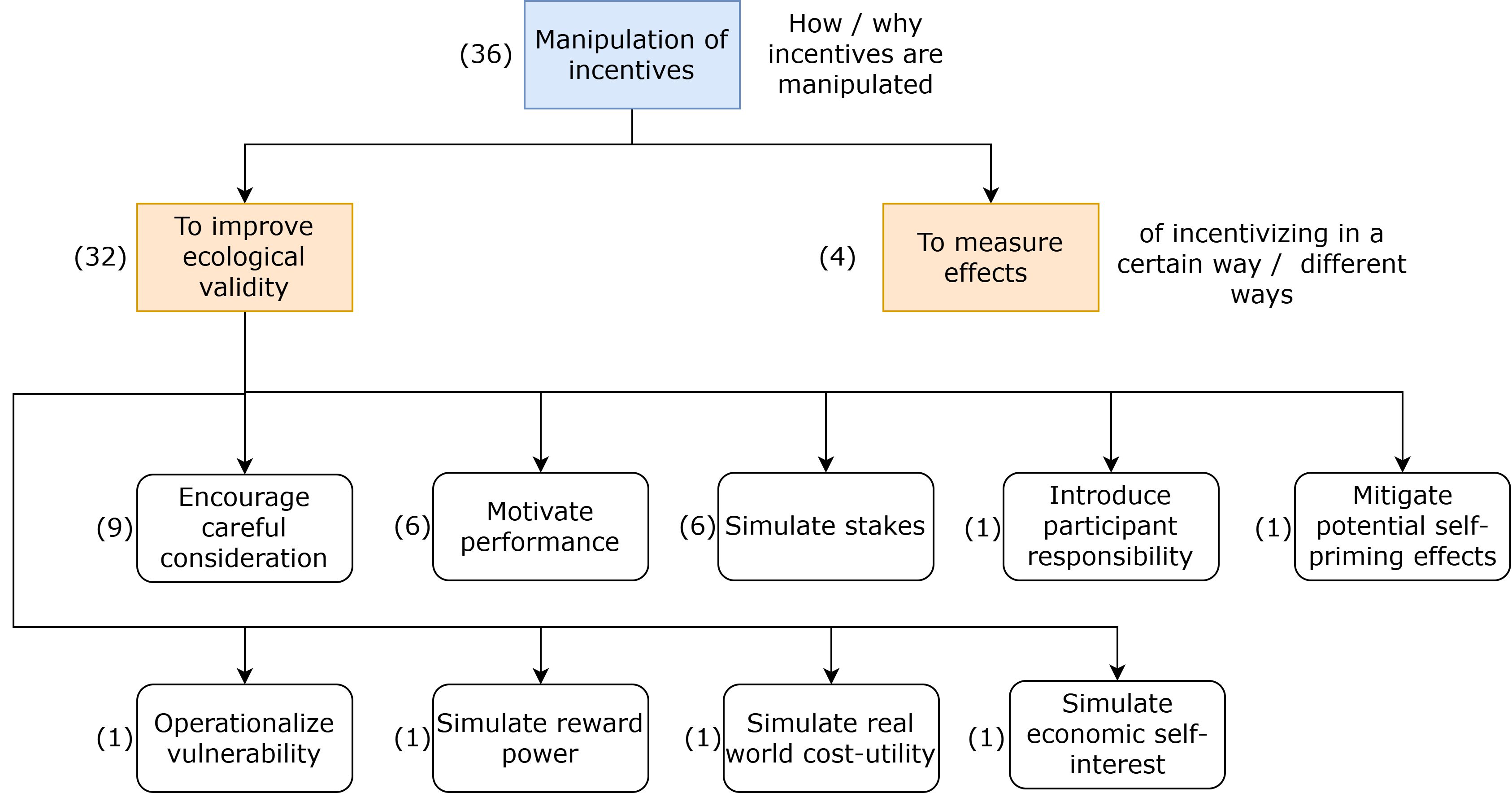}
    \caption{Thematic chart for Theme 2: Manipulation of incentives}
    \label{fig:themes2}
\end{figure}

The next identified theme is that incentives are \textit{manipulated} for various purposes. 
Figure \ref{fig:themes2} illustrates the chart for this theme and its sub-themes.

36/97 papers mention using, or manipulating, incentives for certain purposes.
Largely, the purpose is identified as \textit{improving the ecological validity} of the experiments. For instance, \cite{ConfidenceAndExplanation2020} said, \textit{"We took two measures to improve the ecological validity. First, the decision performance was linked to monetary bonus..."}.

Researchers mention using incentives (mainly bonuses) to: "motivate performance" \cite{ToTrustOrToThink21}, "encourage participants to pay attention" \cite{DresselFarid2018}, "simulate stakes" \cite{21}, "simulate reward power" \cite{81}, and more.
We identify these as examples (and not sub-themes) of the broader theme of improving the ecological validity.


Finally, bonus schemes are also intentionally varied (such as paying high vs. low bonuses) to \textit{measure the effects} of the different schemes on participant performance or research outcomes. 
For instance, \cite{AccuracyOnTrust19} mentioned, \textit{"We also posited and pre-registered two additional hypotheses: [H3] The amount at stake has a significant effect on people’s trust in a model before seeing the  feedback screen. [H4] The amount at stake has a significant effect on people’s trust in a model after seeing the feedback screen [...] to test whether the effect of stated accuracy on trust varies when people have more “skin in the game”." }

{This theme suggests that researchers are aware of the potential limitations of laboratory settings and seek to create more realistic experimental conditions. However, this practice also raises questions about the generalizability of findings to real-world scenarios.}

\subsection{Theme 3: Impact of incentives}
\begin{figure}[!h]
    \centering
    \includegraphics[width=0.8\textwidth]{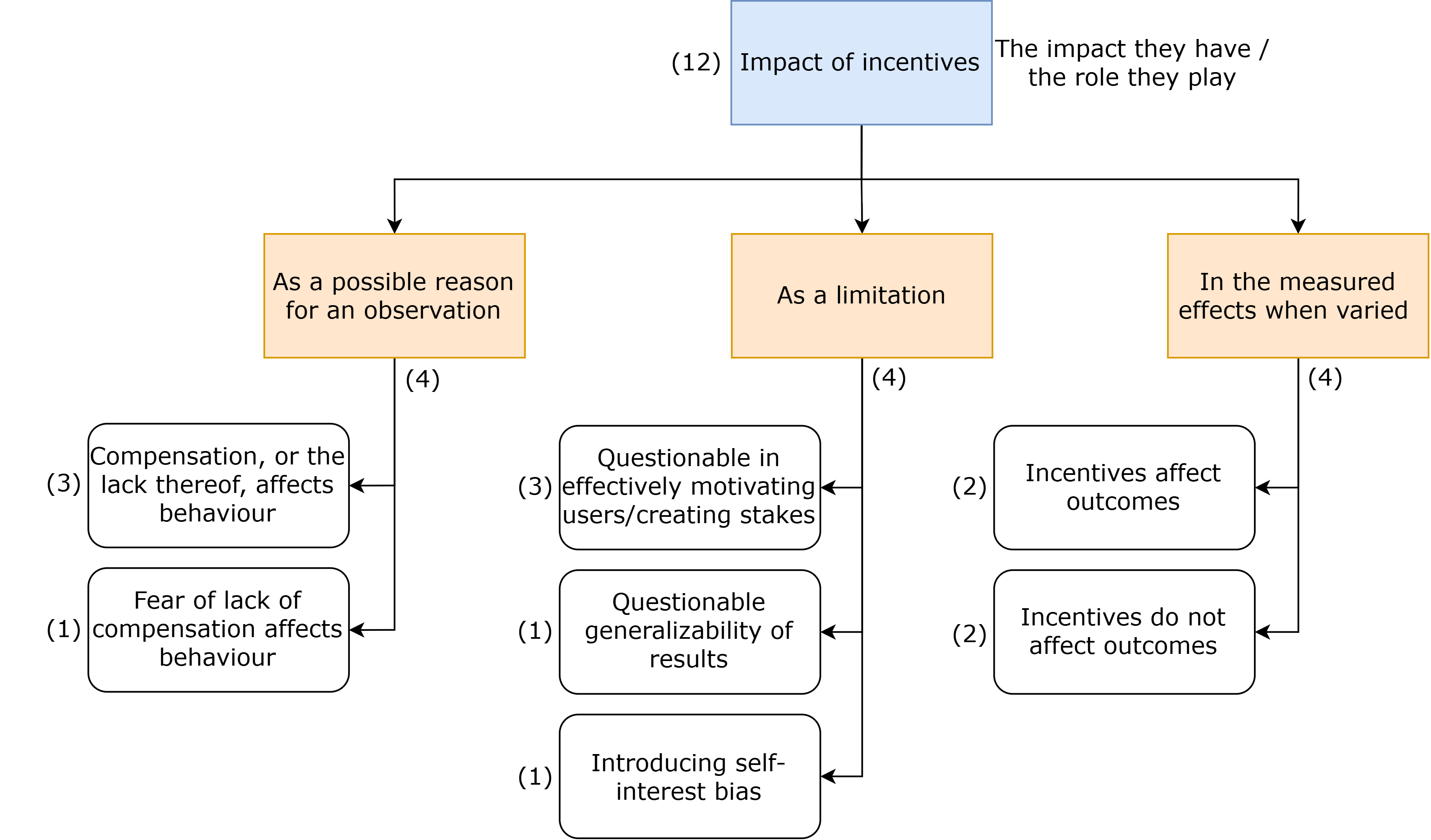}
    \caption{Thematic chart for Theme 3: Impact of incentives}
    \label{fig:themes3}
\end{figure}

The third theme is identified as the \textit{impact} of incentive schemes on the results of a study. 
Figure \ref{fig:themes3} illustrates the chart for this theme and its sub-themes. 

12/97 papers mention the impact of incentives, reflecting on the potential role that incentives played in their studies. 
Some researchers discuss the limitations of incentive schemes, questioning whether they can effectively replicate real-life scenarios or motivate crowd workers, acknowledging uncertainty regarding the generalizability of their results.  
For instance, \cite{overcomingAA2018} noted that, \textit{"Because real life end-users can have very different demographics characteristics and non-monetary incentives and operate in higher-stake environments, we cannot reliably generalize the finding to real workplaces..."}.

In a few cases, incentives are also attributed to as the \textit{potential reason behind an observation}, such as a particular trend being observed because participants may or may not have been motivated to perform due to compensation, or the lack thereof. For example, \cite{55} note that, \textit{"There are two important caveats to the analysis. First, it relies on non-incentivised self-reported data near the end of the experiment. Thus, we cannot verify that subjects reflected carefully on their answers..."}.

Lastly, the few papers that intentionally manipulate incentives to measure their effects also describe their results, leading to a discussion on the impact of incentives. For these studies, different researchers found that incentives may or may not affect performance or outcomes under different conditions.
For instance, \cite{ExplanationsCanReduceOverreliance23} described that, \textit{"...we found an impact of rewards on overreliance..."}. Meanwhile \cite{AccuracyOnTrust19} found that, \textit{"...the amount at stake does not have an effect on laypeople’s trust in a model, at least for the limited range of stakes used in our experiment."}
We note that such studies were conducted for dissimilar tasks and domains, under differing conditions, measuring the effects of rewards on different variables.

{The relatively low number of studies (12/97) explicitly discussing the impact of incentives shows that researchers often omit an investigation of the causal effects of different incentive schemes. This further suggests that the field lacks a comprehensive understanding of how incentives influence participant behavior and research outcomes.}

\subsection{Theme 4: Communication of incentives}
\begin{figure}[!h]
    \centering
    \includegraphics[width=0.5\textwidth]{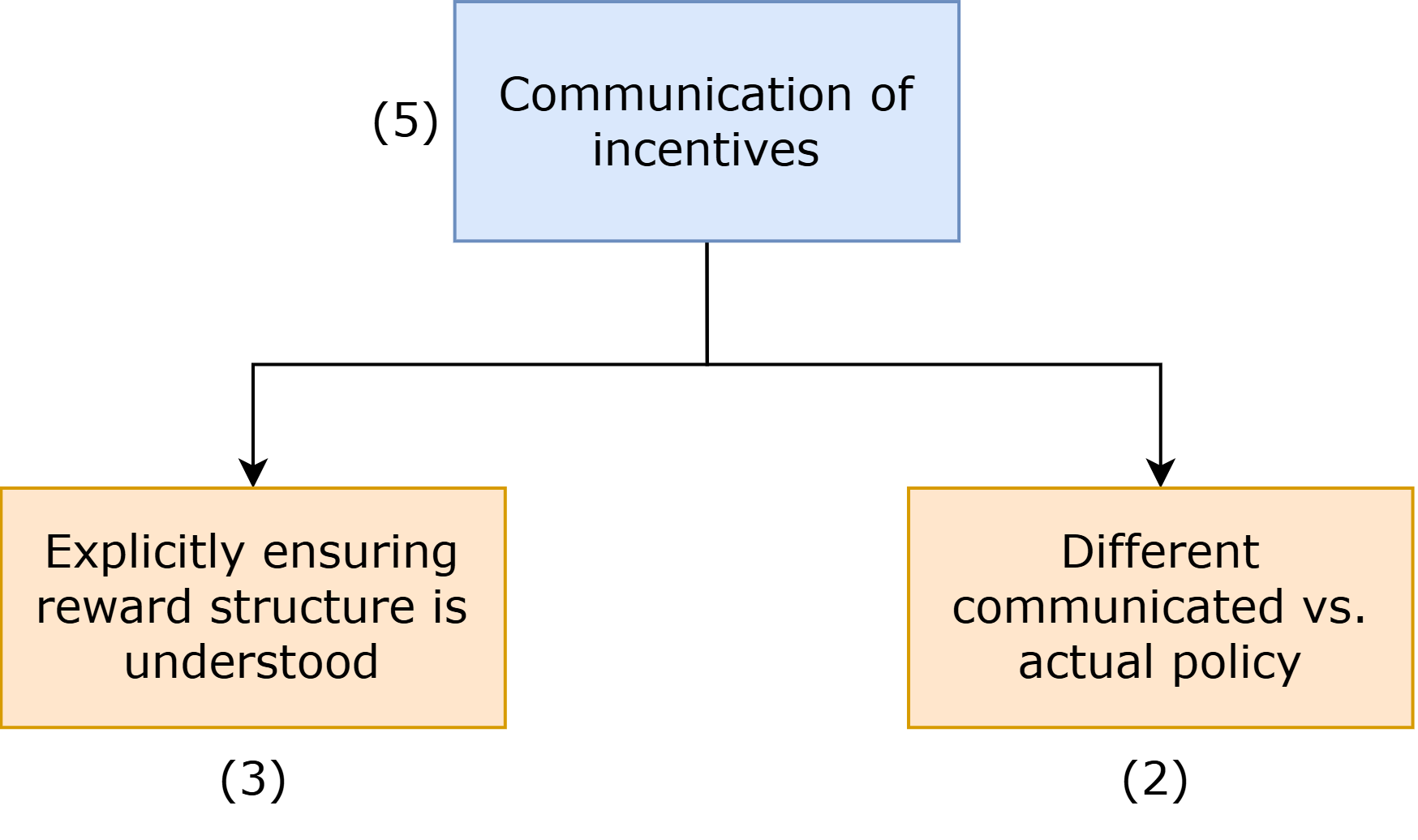}
    \caption{Thematic chart for Theme 4: Communication of incentives to participants}
    \label{fig:themes4}
\end{figure}

This theme highlights the different trends in the \textit{communication} of incentive schemes to participants. Figure \ref{fig:themes4} illustrates the chart for this theme and its sub-themes. 

5/97 papers discussed the communication of their incentive schemes.
In some cases, researchers explicitly ensure that participants understand the pay structure before they begin the task. For example, \cite{19} described that, \textit{"...participants were incentivised to report their true estimates of risk. We articulated this to participants during the tutorial and included a question about the reward structure in the comprehension test to ensure that they understood."}

There are also a couple of instances where the bonus scheme communicated to participants differs from the actual method used to calculate the final payment received by participants. For instance, \cite{38} mentioned that, \textit{"...participants were told that at least the top 50\% of participants would be given a \$2 bonus based on the thoroughness of their evaluations; unbeknownst to them, all ultimately received the bonus."}

{The limited discussion of incentive communication (5/97) indicates a potential gap in the research process, or its reporting. Researchers may not be adequately considering the importance of transparently communicating incentive schemes to participants, or {it is} also possible that the gap lies in the reporting of incentive communication rather than its implementation. Additionally, this highlights that the design of the communication strategy could itself form an integral part of the overall incentive design process.}


\subsection{Theme 5: No mention of incentives}
8/97 papers did not mention anything regarding incentive schemes \cite{PerceptionsofJustice18, AIModeratedDecisionMaking22}.
{The absence of explicit discussions about incentives in some studies (8/97) raises concerns about the potential impact of implicit or unacknowledged incentives on participant behavior and research outcomes, suggesting a lack of transparency in experimental design and reporting practices.}

%% file: 4_Discussion.tex
\section{Discussion}
\label{sec:disc}

Here, we discuss the key findings of the thematic analysis. We reflect on the implications of these observations, raising questions that can guide further investigation.
\subsection{Key Findings}
\subsubsection{Pay Amounts}
The thematic analysis reveals that the current strategies for determining base pay and bonus amounts are predominantly heuristic in nature. Furthermore, a lack of consistency was observed within the calculation and reporting of base pay and bonuses.

For base pay, researchers employed a disparate approach, with some specifying a flat amount and others opting for a pay rate. Additionally, the explanations behind the base pay calculation varied. While some studies explicitly outlined parts of the process, such as estimating task completion time for amount calculation, others lacked such transparency. For bonuses, there was rarely any discussion regarding how the allocated amount was determined.
These observations highlight the need for a more standardised approach towards determining base pay and bonus amounts as well as reporting them. \textbf{\textit{Is there an optimal or appropriate pay amount? Can there be a standardised way of determining it?}}
Could future work explore data-driven methods for establishing optimal pay levels? Can we develop a standardised methodology for calculating base pay and bonus amounts across different tasks and platforms?

{A similar problem has been studied in prior research, as the "Budgeted Task Scheduling" problem. Researchers have proposed algorithms for task allocation and inference schemes, optimized to achieve
the best possible quality of results within a fixed budget \cite{or11, ICML13, budgetfixaamas14}. However, these algorithms have not yet seen widespread adoption among researchers. Such approaches could serve as a foundational basis for researchers to devise their own schemes or for future research to develop more sophisticated algorithms tailored to the specific needs of human-AI decision-making tasks.}

\subsubsection{Aspects of Bonus Schemes}
The examination of bonus schemes further revealed many different approaches within the different aspects of bonuses, such as a variety of types of bonuses (performance-based, completion-based, luck-based), bonus calculation policies (fixed / proportional), and performance evaluation metrics. Notably, not all studies comprehensively addressed each aspect that we identified. Future work should aim at making the process of bonus design more comprehensive. Further investigation is needed into which approaches are most suitable under different task conditions.
    \textbf{\textit{Can we develop a standardised methodology that comprehensively recommends specific bonus calculation methods based on study conditions?}}
Is one type of bonus scheme more appropriate or suitable than the other? Which metrics are most suitable for evaluating performance? What is the best way to map performance to pay? Can these policies be optimised? Under what conditions?

{The different policies identified through the thematic analysis could serve as a starting point for researchers to further refine and experiment with various incentive schemes. By conducting pilot studies and controlled experiments, researchers could investigate the effects of different policies. More sophisticated algorithms along the lines of the ones mentioned in the previous section could also be devised to optimize incentive allocation and maximize participant engagement.}

\subsubsection{Use of Rewards and Penalties}
The analysis highlighted the use of bonuses alongside base pay, with some studies even employing penalties to simulate high stakes. Further research is needed to determine the effectiveness of these approaches.
    \textbf{\textit{Is there value in the use of rewards and penalties? Does it create “higher” stakes?}}
Does the inclusion of bonuses and penalties truly create "higher stakes", or does it introduce ethical concerns? Is there a balance to be struck between positive and negative incentives that optimizes participant motivation without compromising ethical considerations?

{Theoretical frameworks could be leveraged to justify such strategies. For instance, prospect theory \cite{prosp79} suggests that people are more sensitive to losses than gains. By incorporating loss aversion into the design of incentive schemes, researchers could potentially increase motivation and engagement. However, further research would still be needed to determine the effectiveness of such approaches.}


\subsubsection{Improving Ecological Validity}
The analysis highlighted concerns about the ability of incentives to create "stakes" and enhance the ecological validity of studies. Future research should explore methods for evaluating the impact of incentive structures on participant behaviour and their ability to enhance ecological validity in actuality.
    \textbf{\textit{Can it be evaluated and/or ensured that incentives have the intended effect on participant performance and behaviour? How?}}
Can some practices be devised to objectively assess whether incentives truly improve the ecological validity of an experiment?

{To assess the effectiveness of incentives on participant behaviour and ecological validity, researchers can employ various strategies, including pre- and post-study comparisons, controlled experiments, and longitudinal analyses to measure shifts in engagement and performance. Objective evaluation can be achieved through behavioural validation studies, comparisons to real-world scenarios, and multi-dimensional scoring systems that account for task strategies and outcomes.}

\subsubsection{Effects of Incentives}
The analysis highlights a complex relationship between incentive schemes and research outcomes in crowdsourced studies. 
    \textbf{\textit{Which kind of incentive schemes affect outcomes? Which do not? Under what conditions?}}
    
We saw that some studies found that incentive schemes affect outcomes, while others did not. This suggests that {there is} no single "formula" for incentive design. The effectiveness of an incentive scheme likely depends on several factors, such as the experimental conditions and design. This begets a deeper exploration within this focus.
    
\subsubsection{Limitations of Incentives}
The analysis highlighted that researchers show concerns regarding the limitations of using incentives in crowdsourced studies, particularly with respect to the generalizability of their findings and potential bias due to crowd worker motivations.
    \textbf{\textit{Can better incentive design practices mitigate the limitations associated with their use? How can incentives be designed to better understand ways in which they affect results?}}
Can practices be devised to \textit{intentionally design} incentives that can overcome limitations?

{Better incentive design practices can potentially help mitigate limitations by aligning incentives with desired outcomes and reducing unintended effects. Intentional design could include aligning incentives to participant motivations, balancing intrinsic and extrinsic rewards, and ensuring fairness. To better understand their impact, rigorous explorations are need into systematically varying incentives and using behavioural metrics and qualitative feedback to assess their effects. Designing incentives to promote meaningful engagement, rather than purely self-serving interests, could enhance the reliability and ecological validity of results while addressing inherent limitations.}

\subsubsection{Communication Strategies}
Explicitly ensuring that incentive schemes are understood along with performance feedback can improve participant understanding and potentially enhance results, but it might also introduce bias or demotivate participants.
    \textbf{\textit{Is there value in different types of communication strategies for incentives?}}
Is it more appropriate to pay for performance as communicated, or pay everyone equally?

{Transparent communication about how incentives are tied to performance could enhance understanding and motivation but must be carefully designed to avoid introducing bias or discouragement. Paying for performance as communicated aligns with expectations and reinforces accountability but may disadvantage lower-performing participants. Alternatively, equal payment fosters fairness and inclusivity but may reduce motivation for high effort. Balancing these approaches requires careful deliberation and context-specific experimentation to identify strategies that optimise engagement while maintaining fairness and trust.}

\subsubsection{Missing Incentive Schemes}
Lastly, we noticed that some researchers chose not to discuss incentive schemes for their studies. 
    \textbf{\textit{Should researchers be encouraged to consistently report on incentive schemes used in crowdsourced studies, regardless of their perceived relevance? If so, how?}}
We note that the absence of description doesn't necessarily mean incentive schemes weren't thought about or employed. Researchers possibly might not have considered mentioning them relevant to their specific goals. {Nevertheless, transparency enables better replication, comparability, and understanding of results. Researchers could describe the type of incentives, communication strategies, and their alignment with study goals as part of their research design. Standardized reporting frameworks or guidelines could facilitate consistent documentation, ensuring that incentives are considered an integral part of study design and analysis.}

\subsection{Reflection}
The questions raised above fall into two main categories: \textit{methodological challenges} and \textit{exploratory avenues}. Methodological challenges concern the practical implementation of incentive schemes, such as the \textit{design} of pay amounts and structures. Exploratory avenues encompass broader research directions, including studying the \textit{effects} of manipulation and different communication strategies.
{We note that the methodological challenges are closely intertwined with the exploratory avenues. For instance, the choice of performance metrics and the design of bonus schemes can impact how we study the effects of incentive design on participant motivation and task performance. Therefore, a deeper understanding of how these factors interact is crucial for designing effective incentive schemes.}

While a comprehensive exploration of all these questions was beyond the scope of this work, we prioritised addressing the methodological challenges surrounding incentive design, as these concerns aligned with our defined research agenda.
The findings of the thematic analysis further solidify the need for standardization in light of the challenges inherent to human-AI decision-making tasks, calling into question the very notion of "appropriate" incentive design.
Thus, in order to address RQ2 and RQ3, we took up the task of establishing a standardised solution that can allow researchers to \textit{tune} "appropriate" incentive schemes for their specific human-AI decision-making studies. 
To address RQ2 and RQ3, we identified that a well-designed {framework} can foster a standardised, systematic, and comprehensive approach that can guide research towards designing and documenting effective incentive schemes. It can additionally act as a safeguard against unintentionally overlooking the important aspects of incentive design. 
{Further, a deeper exploration into the other aspects, such as the effects of incentives on research outcomes and the ethical implications of high-stakes designs, should be combined with our framework in the future.}

%% file: 5_Framework.tex
\section{The Incentive-Tuning Framework}
\label{sec:framework}


\begin{framed}
   \begin{enumerate}
    \item \textbf{Identifying the purpose of employing an incentive scheme}: \textit{What is the goal we wish to achieve by incentivizing participants? What are the desired behaviours we expect from participants?}

    \item \textbf{Coming up with a base pay}: \textit{What is the {baseline} rate participants must be paid to ensure the identified goals are met {while ensuring fair pay}? Which characteristics of the task can affect this amount?}
    
    \item \textbf{Designing a bonus structure}: \textit{Can offering bonuses help reach the identified goals?}
    \begin{enumerate}
        \item \textbf{Coming up with a bonus amount}: \textit{How much total amount should be allocated for bonuses?}
        \item \textbf{Deciding the type of bonus(es)}: \textit{What type of bonus is the most suitable for reaching the identified goal?}
        \\In case of performance-based bonuses:
        \item \textbf{Deciding the performance evaluation metrics}: \textit{How to evaluate participant performance? What is the behaviour that we wish to reward?}
        \item \textbf{Deciding the policy for mapping rewards to performance}: \textit{How to map performance to rewards?}
    \end{enumerate}
    \item \textbf{Gathering participant feedback}: \textit{What do we wish to understand from the participants' perspective? Can we augment the experiment pipeline to gather such feedback?}
    \item \textbf{Reflecting on design implications}: \textit{Did the design achieve the desired goals? Were there any unintended consequences?}
\end{enumerate} 
\end{framed}

\subsection{Step 1: Identifying The Purpose}
Explicitly identifying the purpose of employing an incentive scheme can facilitate \textit{intentional design}, ensuring that it aligns with the research goals. Possible goals might include meeting ethical standards, improving the ecological validity \cite{Ecovald2018} of the study, creating or simulating stakes, motivating participant performance, or encouraging desired behaviours beyond task completion or performance. {During this step, researchers must recognize that ensuring fair compensation is their ethical obligation, and identify participant welfare as part of their goals.}

\subsection{Step 2: Coming Up With A Base Pay}
The identified goals as well as the characteristics of the task, such as the task complexity and required expertise, can affect the base pay amount. For example, complex tasks with higher cognitive loads could discourage participation due to perceived low value for time invested \cite{CognLoad2021}. Thus, such tasks may warrant higher pay to attract participants and motivate them to perform well. Similarly, tasks requiring specialised skills or in-depth knowledge typically warrant a higher base pay compared to tasks that a layperson can do with minimal training \cite{LabourLawsCS2011}.  

Other factors such as the minimum wage and platform recommended rates can guide the ethical standard for pay. Pricing of similar past studies can also be relevant as prior work has shown that there can be an anchoring effect \cite{Incentives2009, Ancheffect2013} on participants' perceptions of fair pay.

Lastly, conducting pilot studies is becoming an increasingly encouraged practice in human-AI research as they provide an opportunity for researchers to identify potential issues with their task design and can help calibrate task parameters for eventually conducting the experiment at a larger scale \cite{ImpofPS2002, StateofPS24}.
Pilot studies have been used to estimate the time taken to complete a task in previous studies. This practice can be extended further, such as to assess the sufficiency of the base pay to be offered, and help tune it to better meet participant expectations of fair pay. {Prior work has proposed rapid prototyping and a data-driven effort metric for measuring the effort required from participants to complete a task~\cite{cheng2015measuring}. As an alternative or a complement to pilot studies, this can be a suitable method to estimate the base pay.}

\subsection{Step 3: Designing A Bonus Structure}
{While a fair base pay is essential and prior work has explicitly addressed this~\cite{whiting2019fair,irani2013turkopticon},} it might not fully capture the nuances of task complexity or the desired level of performance. Bonuses provide a way to incentivize crowd workers and encourage specific behaviours, such as performance, that go beyond simply completing the task.
Specially in \textit{high-risk} scenarios, bonuses can introduce a sense of raised stakes for the participants who are often primarily driven by monetary goals \cite{MonetaryGoalsCS2015}.
At the same time, we acknowledge that bonuses require nuanced consideration. A strategy such as simply offering large bonuses, may not directly translate to higher quality work \cite{Incentives2009, HigherInc19}. It is thus important to strike a balance between the compensation amount and its distribution, while ensuring alignment with the specific research goals. Hence, we suggest that researchers should critically evaluate the behaviours they aim to incentivize when designing bonus structures. 
We discuss the different aspects of a bonus scheme below.

\subsubsection*{Coming up with a bonus amount}
As with base pay, past pricing for similar tasks can serve as the starting point for the maximum attainable bonus amount and pilot studies can be used to refine it, for example by including questions about participant satisfaction and perceptions of fairness regarding the additional rewards. 

Further, we identified three possible types of bonus schemes through the thematic analysis: performance-based, completion-based, or randomised.
Randomised bonuses have a potential trade-off between possibly discouraging high-performing participants or encouraging low-performing participants to engage more deeply, and should thus be used cautiously. Completion-based bonuses can be helpful for encouraging participation in optional tasks (such as surveys), where the main goal is to gather a sufficient amount of data. Finally, for performance-based bonuses, which are the most commonly used to motivate desired behaviours, their effectiveness relies on choosing the right performance evaluation metrics as well as the policy for mapping performance to rewards. Performance could mean accuracy or speed, but the chosen metric should truly reflects task goals and not encourage participants to act in an undesirably biased way.
Further, performance could also mean more than simply a metric - it could mean engaging in any kind of desired behaviour.
For example, if explaining reasoning behind decisions is important, offer bonuses for in-depth explanations. If reading the provided information carefully is deemed important, incentivize attention or comprehension checks.
Task characteristics such as decision subjectivity can also affect these factors. For example, if decision subjectivity is high, open-ended questions could be used to assess performance. 
We also identified that some past studies made use of penalties to create high stakes. This can potentially be effective by appealing to the loss averse tendencies of participants \cite{Decisionsunderrisk1979}. However, at the same time, there is a potential trade-off that excessive penalties might discourage participation. Thus, caution must be exercised while incorporating them into the design.
Further, there can be several ways to map performance to pay, such as: rewarding over a threshold performance or increasingly rewarding with better performance. This requires identifying what works best for the task considering the research goals. Delving deeper into incentive design research to identify possibly suitable methods of designing optimal policies may prove to be beneficial. 
Finally, pilot studies can again be leveraged to see how participants respond to different bonus schemes.

\subsection{Step 4: Gathering Participant Feedback}
Post-task or exit surveys can be useful tools for understanding participant experiences and motivations. They can include questions about perceived fairness of pay and participant motivation, also allowing researchers to compare responses and assess if the pilot study findings regarding similar questions hold true in the main experiment.
For example, if the results show a less than expected accuracy rate, survey responses about perceived unfair pay might indicate that participants were less motivated to put in their best effort. This information can be crucial for interpreting the results as well as their generalizability.

Ultimately, this practice can allow us to assess the effectiveness of the chosen incentive scheme. By asking participants questions such as whether they felt the pay was fair and if the bonus structure motivated them, researchers can gauge whether the incentive scheme achieved its goals, such as attracting qualified participants or encouraging desired behaviours. Further, it can aid the next step of reflecting on the design implications by providing the participants' perspective.
Such practices could help refine incentive schemes for future research. For example, if participants report that they felt inadequately compensated or unmotivated by the bonus structure, it could be adjusted to better meet participants' expectations for future studies.

\subsection{Step 5: Reflecting On Design Implications}
So far, we have suggested that researchers should engage in \textit{intentional design}. We encouraged researchers to understand trade-offs, justify their decisions, and gather feedback from participants, in order to eventually be able to reflect on the implications of their design.\\
To bring the design process to fruition, the final step of the {framework} prompts researchers to also reflect on the potential effects of their design on their research outcomes.
By doing so, researchers can be more confident in their findings and provide valuable insights into the field. Additionally, it fosters a culture of transparency and accountability, allowing for making improvements upon existing methods for future research.

{By outlining the above steps, the framework aims to prompt researchers to explicitly consider and define their research goals, including those that may seem less tangible, and approach the incentive design process in a systematic manner, highlighting its key aspects. By doing so, the framework can help ensure that incentive schemes are designed to directly support the specific research objectives and can be evaluated effectively. Ultimately, how researchers operationalize the framework may vary depending on the context of the research, but the core principle remains the same: the incentive design should reflect the goals that researchers set out to achieve.}

\subsection{Documenting the Design of Incentive Schemes}
While the Incentive-Tuning {Framework} provides a systematic and comprehensive overview of items researchers must report, some ambiguities remain regarding the format and data presentation for reporting each element.
For example, the thematic analysis highlighted inconsistencies in how base pay is reported, including using "fixed amount" vs. "pay rate," and mentioning or omitting the resulting total pay. 
Another area of ambiguity identified is the use of averages vs. medians to report resultant compensation. While both offer valid summaries of data, a consistent approach across studies is vital.
 
\subsubsection{A Template}\label{Ch4:Template}
A standardised reporting \textit{template}, built upon the foundation provided by the items of the Incentive-Tuning {Framework}, can potentially address the challenges identified in the previous section. 

The following template can be used to capture details regarding the incentive scheme in the experimental design section of a research article:
\begin{framed}
    The purpose of our incentive design was to ensure [identify primary goals].
    Participants received a resultant pay of [average and median resultant pay amount] based on a base pay rate of [base pay per hour] and [average and median bonus payout].
    
    The base pay was set as [amount] for completing the task to ensure fair compensation, considering [rationale, e.g., platform recommendations / minimum wage / past pricing / pilot study feedback / specific task characteristic].
    [Can elaborate further as per choice].
    
    To further [incentivize / ensure / motivate] the [identified goals] a [performance]-based bonus structure was implemented. [Decisions due to specific task characteristics (e.g. use of penalties because of high perceived risk].
    Maximum bonus payout was set as [amount]. Participant performance was evaluated based on [evaluation metrics (e.g. accuracy)], calculated as [reward mapping policy]. 
    [Can elaborate further as per choice].

    [Optional] Participants received additional bonuses for [task specific considerations] to encourage [desired behaviours].

    [Optional] More details, survey feedback etc.
\end{framed}

\textit{Note: It is important to acknowledge that the suggested phrases within the template are not intended to be rigidly adhered to. Their primary function is to illustrate the recommended structure for reporting incentive design decisions. Researchers can adapt these phrases to fit the specific context of their study while maintaining overall clarity and consistency in reporting.}

We notice that the template only addresses the first three items of the {framework}. Given the highly context-dependent nature of the descriptions and discussions pertaining to the remaining items, their completion is left to the discretion of the researchers. To this end, we make the following recommendations:

The study design should explicitly describe the methods employed to collect feedback during the experiment (e.g., surveys, open-ended questions, interviews). In the results or observations section, researchers should present the obtained feedback data.

The discussion section should include reflections on the design choices and feedback. This may encompass insights gained from the feedback data, its influence on the overall research findings, and potential areas for improvement in future studies.

\subsubsection{A Repository}\label{Ch4:Repository}
The dataset that was built for conducting the literature review is a significant corpus capturing incentive design in existing literature. We identified that by collating and presenting relevant items from within this dataset in a public repository, we could create a valuable resource for future researchers.
Hence, we established a public repository on GitHub\footnote{\url{www.github.com}} to promote transparency and collaboration within incentive design for human-AI decision-making studies.

\begin{framed}
    The repository can be accessed here to view the source code or raise pull requests:
    \href{https://github.com/simrankaur1509/IncentiveSchemesForHAIDMstudies}{GitHub Repository}.
\end{framed}

\begin{framed}
    The tabulated incentive scheme data from published articles compiled so far can directly be viewed here: 
 \href{https://simrankaur1509.github.io/IncentiveSchemesForHAIDMstudies/}{GitHub Pages}.
\end{framed}

This repository is aimed to serve as a central hub for researchers to share and access incentive design knowledge from past published research. It is open for public access, currently being actively populated with the design decisions extracted from prior work. We also invite researchers to contribute the incentive schemes they design in the future to this repository.

\subsection{{Case Studies: Applying the Incentive Tuning framework}}

{We present two case studies to demonstrate the application of our proposed framework. To this end, we selected two research papers after obtaining consent from the corresponding authors.}

\subsection*{{Case Study I}}
{For the first case study, we considered the paper titled, "How do you
 feel? Measuring User-Perceived Value for Rejecting Machine Decisions in
 Hate Speech Detection" by \cite{lammerts2023you}. It mentions a simple
 incentive scheme:}
 \begin{framed}
      {"Every participant is paid an hourly wage of 9 GBP, exceeding the
 UK minimum wage at the time of the study."}
 \end{framed}

 {After taking a closer look at the experiment design and study goals
 as described in the paper, we present our findings on designing
 the incentive scheme for this study through application of our framework.}

\begin{itemize}
    \item  {\textbf{Step 1: Identifying the purpose}: We identify that the authors
 wish for the participants to thoroughly process information and focus on
 harm evaluation. Additionally, we surmise that the authors have the
 goals of providing fair compensation, enhancing ecological validity, and
 simulating real-world stakes.}
    \item {\textbf{Step 2: Coming up with a base pay}: The chosen
 base pay of £9 per hour exceeds the minimum wage, addressing fair
 compensation. In trying to explore past-pricing for studies in the toxicity
 classification domain, we did not find many relevant articles addressing
  similar tasks. Since no specific skills are required, the base pay
 seems appropriate. However, considering the high number of tasks (40)
 per participant, we must consider strategies for aiding participant
 motivation throughout and compensating them for the time invested. A
 simple strategy would be increasing the base pay itself. We note that a
 pilot study was conducted by the authors. We recommend enhancing it
 to gauge engagement and satisfaction by including questions such as,
 "Did you feel the need to take any breaks while performing the tasks?
 If so, how many and for how long?" and "Did you feel the pay was fair
 compensation for the time and effort required?"}
 \item {\textbf{Step 3: Designing the bonus structure}: We note that the
 authors did not offer bonuses. Offering bonuses can aid the goal
 of improving ecological validity, simulating stakes and encouraging
 desired behaviours in participants. We recommend that offering small
 performance-based bonuses could be effective.}
 
 {The authors chose not to incentivize correct answers and indicated
 that their primary goal is not to encourage accuracy, but evaluating
 user perceptions of value. They included "lengthy
 descriptions" of the task instead of rewards to direct participants’ focus
 towards evaluating harm, and included attention checks to filter out
 inattentive participants.}
 {To better pursue this goal, we suggest including comprehension
 checks that focus on processing task information and consider accuracy
 on them as the evaluation metric for performance-based bonuses.
 Rewarding participants who pass these checks with a flat bonus can
 encourage attentiveness given the "lengthy descriptions" and work
 within budget constraints. Thus, the performance metric in this
 case would be comprehension check accuracy and the reward mapping
 policy would be 100\% reward on 100\% accuracy.}
 
 {Further, we note that the study deals with a high-risk scenario (hate
 speech classification). As the authors highlight the importance of users
 understanding the consequences of incorrect decisions, levying small
 penalties emerges as an option. However, to combat participants
 potentially getting discouraged, there should be sufficient positive
 rewards as well. But we identified that encouraging task accuracy was not
 a goal, hence decided not to suggest task accuracy-based rewards. Thus,
 we conclude not to use penalties.}
 \item {\textbf{Step 4: Gathering participant feedback}: Especially since we
 did not take any explicit steps to create high stakes by introducing
 consequences (such as through penalties), we recommend using post-task surveys to get a better understanding of whether participants
 understood the consequences and stakes of the decision-making
 scenario. This is crucial, particularly if budget constraints limit
 implementing all suggestions. Understanding participant perspectives
 would help in interpreting the results.}
 \item {\textbf{Step 5: Reflecting on design implications}: The current design
 prioritizes fairness but lacks sufficient strategies to maintain participant
 motivation throughout the lengthy task list. Additionally, the absence
 of consequences might lead to underestimating the importance of
 evaluating harm. While we cannot truly gauge the outcomes of the study
 in this regard unless replicated, potential inconsistencies in how
 participants evaluate harm could be attributed to such factors.}
\end{itemize}

\subsection*{{Case Study II}}
{For the second case study we considered the paper titled, "Dealing with Uncertainty: Understanding the Impact of Prognostic Versus Diagnostic Tasks on Trust and Reliance in Human-AI Decision-Making" by \cite{Uncertainty24}.
This work describes the incentive scheme in more detail:}

\begin{framed}
        {"All participants were compensated at the fixed rate of 8 GBP per hour regardless of their performance in the study. Additionally, participants received bonus rewards amounting to 0.2 GBP for each accurate response they provided during the study period. Overall, participants earned an average of 8.44 GBP per hour, well over the wage considered to be ‘good’ and recommended by the Prolific platform."}
\end{framed}

{
\textbf{Step 1: Identifying the purpose}:
We identify the purpose as ensuring fair compensation for participants, enhancing ecological validity of the experiment, encouraging accurate decision-making in trip-planning tasks, and simulating real-world decision-making.}

{\textbf{Step 2: Coming up with a base pay}:
The chosen base pay meets \textit{platform recommendations} and there are not enough \textit{prior studies} using the same task domain or structure for additional insight.}

{We note that the authors manipulate the complexity of tasks by adjusting the number of features and constraints presented to participants. Considering the complexity levels (low, medium, high) and the cognitive load induced by the tasks, we consider adjusting the base pay based on \textit{task complexity}. Possible calibrated base pay could be the same as described by the authors for low complexity tasks, slightly higher for medium complexity tasks, and significantly higher for high complexity tasks. 
Appropriately incentivizing participants and addressing the varying task complexity levels through incentives can help the researchers reflect on the effect of participant motivation on performance while analyzing their results. For example, the authors make an observation regarding the decline in performance for medium/high complexity tasks. This could also be attributed to lower participant motivation due to increased cognitive effort. If researchers account for the additional cognitive effort while incentivizing participants, they can assert that they took intentional measures to combat this and thus they can attribute the decline in performance to other factors (such as uncertainty or complexity itself) with more confidence, hence improving the ecological validity of their findings.}
{At the same time, we notice a \textit{trade-off} in implementing such a pay scheme. Paying participants differently based on the tasks they receive can be perceived as unfair; it can be argued that all participants deserve equal pay for their time, regardless of task complexity.}

{The authors did not mention a pilot study.
A pilot study with targeted feedback questions could be used to validate 
the different approaches, gauging participant motivation and perceived fairness with different pay structures. A pilot could also help assess the actual time and effort required for different complexity levels. This can help in objectively calibrating the pay scale.
Further, we could include cognitive effort questionnaires as well as open-ended questions such as "Were you motivated to put in genuine effort to perform the task?" and "Was the reward fair and satisfactory as per your expectations?" Based on the responses, the amount and structure can be further tweaked.} 


{\textbf{Step 3: Designing the bonus structure}:
The authors used \textit{performance-based bonuses}, with the evaluation metric as \textit{accuracy} to encourage accurate trip-planning, addressing the \textit{study goal} of encouraging correct decision-making. Also, since \textit{decision subjectivity} is low, accuracy could be a suitable metric to consider. }
%
{An 
alternative design could be rewarding for \textit{accuracy when the AI is wrong}, incentivizing genuine effort and discouraging blind reliance on the AI. This could re-focus participants on the goal of accurate trip-planning instead of maximizing their rewards. Another option would be to explore rewarding for "accuracy-wid" (final correct decision with initial disagreement with AI) or appropriate reliance. While directly rewarding appropriate reliance might influence behaviour, it may also lead to more confident interpretations of results.}

{Ultimately, we have a \textit{trade-off} between two choices for the \textit{performance evaluation metric}. When making a decision, we acknowledge the implications of each choice:}
\begin{enumerate}[nosep]
    \item {\textit{Encourage appropriate reliance, use accuracy-wid or reward higher bonuses for correct answers when the AI is wrong}: The implication of this choice would be that we are encouraging what we wish to measure, as the researchers' goal is to measure appropriate reliance itself.}
    \item {\textit{Encourage overall accuracy without influencing reliance}: Here, we need to acknowledge the possibility for overreliance on the AI.}
\end{enumerate}

{
However, we note that researchers wish to measure appropriate reliance \textit{in the wild}. Thus, we conclude that \textit{we should not encourage it by rewarding it}. }
%
{As for \textit{mapping rewards to performance}, increasing rewards with increasing performance could be a strategy to further motivate participants to make correct decisions, if the budget permits.}
{Finally, we observe that the perceived risk is relatively low, so we do not consider the use of negative rewards like penalties.}

{\textbf{Step 4: Gathering participant feedback}:
We recommend including questions about participant motivation, cognitive load, and perceptions of pay fairness in the post-task survey as they remain crucial for understanding participant perspectives and interpreting the results.}

{\textbf{Step 5: Reflecting on design implications}: The current incentive scheme prioritizes fairness and encourages accurate decision-making. However, the potential for overreliance on AI can have implications on the ecological validity and must be acknowledged explicitly. We discussed the implications when choosing each of the different evaluation metrics in detail in Step 3. Such discussions should be included when presenting the results of the study.}

{\textit{We acknowledge that our recommendations and final decisions may not present the sole appropriate design choices. The primary goal of presenting the case studies is to highlight that through the Incentive-Tuning Framework we can stimulate a broader discussion on each aspect of the incentive scheme. We aimed to demonstrate how researchers can carefully consider trade-offs and justify their choices when making decisions, guided by the framework.}}

%% file: 6_Implications.tex
\section{Implications}
\label{sec:implications}

\subsection{Implications for methodology and for the HCI community}
Our work has important implications for the methods used to design and analyze empirical human-AI decision-making studies. 
The Incentive-Tuning {Framework} proposed by us offers a valuable foundation for researchers to design effective and \textit{intentional} incentive schemes. By systematically encouraging consideration for the study's goals, incentive design aspects, participant behaviour, and potential trade-offs, the {framework} empowers researchers to design incentive schemes that promote desired behaviours. It remains extensible and can in the future be updated to incorporate further advancements that might happen in the field.
The standardised approach that the {framework} provides to incentive design not only has the \textit{potential to improve data quality} but it can also \textit{facilitate comparisons} across different research projects, fostering collaboration and accelerating scientific progress in human-AI decision-making.
At the same time, the {framework} acknowledges that a "one-size-fits-all" approach might not be optimal when designing incentives for every unique study.
It offers guidance to researchers while allowing for customization based on specific study requirements and research questions. This, in turn, will contribute to more \textit{robust} human-AI decision-making research, paving the way for advancements in this rapidly evolving field.

Further, the template we provide for documenting incentive schemes offers a standardised format, which has practical implications. It can ensure that researchers capture all the relevant information following the application of the {framework}. This would make it easier for readers to understand the intent with which incentives were designed and their potential influence on participant behaviour and study findings.
The public-access online repository we set up also allows researchers to access and share detailed information about incentive schemes, including the rationales behind design decisions. The collaborative aspect of the repository offers several benefits. It facilitates \textit{knowledge sharing} as researchers can easily find existing incentive schemes relevant to their work, making it possible to learn from each other's incentive design choices, accelerating progress in the field.
It also fosters \textit{transparency} by allowing the community to review the incentive design decisions for different studies.

\subsection{Implications for theory}
The thematic analysis we conducted provides valuable insights into the various incentive design approaches employed by researchers and the diverse discussions surrounding the role and impact of incentives.
Our exploration also revealed unanswered questions that demand further investigation, like incentive manipulation and communication. 

The overview we provide can be a \textit{solid foundation} for researchers to delve deeper into specific aspects of incentive design. Our insights can be a powerful lens for researchers to look into a topic, and use them to kick-start or shape their own research and further contribute to the field's advancement. By delving deeper into the still unanswered questions, researchers could refine the approach further to create more effective incentive schemes and contribute to the theoretical underpinnings of our current understanding of incentive design.
{Future research could experiment with different incentive structures to directly compare their impact on participant behavior and data quality for human-AI decision-making tasks. This can be extended to include longitudinal studies~\cite{soprano2024longitudinal} that track participant behavior over time to understand the long-term effects of different incentive schemes as well as qualitative studies to gather insights into participants' perceptions of incentives and their impact on motivation and engagement. 
Additionally, beyond targeted studies focused solely on incentive design, researchers could integrate hypotheses about the impact of incentive schemes into their studies on human-AI decision-making that investigate other factors, such as trust, reliance, or fairness.}

{Pursuing these directions can lead to a more comprehensive understanding of the complex interplay between incentives, participant behavior, and research outcomes in human-AI decision-making. This knowledge can be leveraged to design more effective and ethical incentive schemes that optimize participant engagement, data quality, and overall research outcomes.}

\subsection{Caveats, Limitations, and Other Considerations}


\subsubsection{{Non-monetary Incentives}}
{Although our work is scoped to the extent of considering monetary incentives in crowdsourced human-AI decision-making studies, it is important to note that there can be a broader spectrum
of alternative incentive designs such as reputation-based~\cite{whiting2017crowd,zhang2012reputation}, gamification-based (e.g., using badges or leaderboards, competitions) ~\cite{rokicki2015groupsourcing,morschheuser2017gamified}, social rewards~\cite{feyisetan2017social,fan2020crowdco,de2024we}, performance feedback~\cite{dow2012shepherding}, or learning incentives~\cite{abbas2022goal,margaryan2020learning}. A wealth of literature in human computation and crowdsourcing has explored the role of such incentives in shaping crowd work~\cite{gadiraju2019crowd,mao2013volunteering,radanovic2016incentives}. Prior work has also demonstrated the effectiveness of incentivizing participants with personalized feedback in exchange for participation in behavioral studies `in-the-wild'~\cite{reinecke2015labinthewild,huber2020conducting}. Future work can explore how alternative non-monetary incentive designs can be effectively leveraged to support the ecologically valid orchestration of human-AI decision-making studies. How exactly such alternatives compare to monetary incentives is an open question worth exploring.}

\subsubsection{Non-exhaustiveness of the Framework and Iterative Improvement}
The research landscape of human-AI decision-making as well as incentive design in crowdsourcing is constantly evolving, and new challenges or unforeseen scenarios might arise. New types of tasks and research questions might emerge, requiring considerations that {have not} been captured in the thematic analysis and {framework} yet. Further, unexpected complexities in the research design, participant behaviour, or platform functionalities might necessitate adapting or going beyond the suggestions we offered.
Acknowledging this inherent limitation, we suggest that we should treat the {framework} as a guide, not a rigid formula. We should be prepared to adapt and refine the suggestions based on the scenarios that may arise.
Further, we believe that the {framework} itself should be a living document. As researchers gain experience using it and encounter new challenges, the {framework} should be iteratively improved to incorporate best practices and address emerging issues.

\subsubsection{Barriers to Adoption and Individual Differences}\label{Ch3:Limitations:Adoption}
It is possible that researchers might perceive the {framework} as complex or time-consuming to {adopt and implement}, especially for smaller studies with limited resources.
{Conducting retrospective analysis or} controlled experiments that investigate the impact of using the {framework} on research outcomes {to shed light on the practical utility and value of using the framework is a meaningful direction of future work}.
%
This could potentially enable researchers {and practitioners deploying human-centered AI studies to appreciate} its benefits and {help them} adopt it in their own work.

The Incentive-Tuning {Framework} offers a structured, systematic approach for designing incentive schemes while allowing flexibility to address different types of studies. However, {it is} important to acknowledge the potential impact of individual differences among the participants of a study. {There is} no single "right" incentive scheme that will universally motivate every participant in the same way.
While the {framework} focuses on a systematic, step-by-step approach, researchers should also be aware of alternative approaches that cater to individual differences. These approaches include dynamic pricing or adaptive incentives \cite{DynamicPricing23, AdaptveIncentives14}. These methods involve adjusting incentive structures based on factors such as participant skill level, performance history, or even real-time task complexity. They utilize algorithms to continuously adapt incentive structures during the experiment, based on participant behaviour and engagement levels. This allows for a more personalised approach to incentivization. While such approaches hold promise, their adoption is still limited due to factors such as increased design complexity and the need for advanced data analytics capabilities.

In the future, the {framework} could be enhanced to incorporate considerations for individual differences. The framework could potentially be expanded to include steps that guide researchers in exploring dynamic pricing or even developing basic adaptive incentive structures within the constraints of their specific study design.

\subsubsection{Potential Biases and Assumptions}
{It is} important to acknowledge potential cognitive biases that can influence the design process the Incentive-Tuning {Framework} outlines. One such bias is the self-interest bias \cite{CBChecklist}. This bias highlights that participants naturally prioritize their own monetary goals and may be inclined to behave or perform in ways that maximize their rewards.
We try to address this bias by highlighting the need for caution when designing performance-based bonuses. Unrestricted bonuses can inadvertently exacerbate self-interest bias. However, we {do not} dismiss bonuses or rewarding for performance. Instead, we acknowledge that self-interest bias can be leveraged to motivate participants. Carefully aligning reward structures with desired behaviours can allow researchers to utilize the self-interest bias to encourage participants to exert effort, focus on desired metrics, and contribute high-quality work.
However, there is always the possibility that rewards cause participants to prioritize maximizing their compensation over providing high-quality data. We encourage researchers to integrate quality control measures to mitigate this risk and ensure the validity of their findings.

Finally, a core assumption of this work lies in the understanding that incentives can influence crowd worker behaviour. While this notion is widely supported \cite{MotivIncent17, Monetary78, MoreThanMoneyKaufman11, CSGhezzi2017}, previous research also highlights instances where incentives may not have a significant effect \cite{HumansForegoRewards23}. We did not delve into the specific conditions under which incentives might be less or more impactful. Therefore, the extent to which the designed incentive schemes would truly influence behaviour cannot be definitively established. Interestingly, while conducting our thematic analysis this very dilemma emerged as a sub-theme, but a deeper exploration of this dynamic fell outside the scope of this work. Future research could explore the nuances of incentive design and their potential effects on study participants under varying conditions.
{We emphasize that incentive design remains a crucial part of any crowdsourced study. The tools developed in this work to aid this process are still valuable, at least in the context of good research practices. By following a rigorous process that aligns study goals with incentives and engaging in the \textit{intentional} design and documentation of incentives, researchers can conduct their due diligence.}

\subsubsection{Ethical Considerations}
The Incentive-Tuning {Framework} prioritizes the design of incentive schemes that are both effective and ethical. However, {it is} crucial to acknowledge that some of our suggestions touch upon ethical considerations that researchers should carefully navigate.

One such ethical concern is exploitation. We highlight the importance of ethical wages within our suggestions. We assert that researchers have a fundamental ethical obligation to ensure fair compensation for participants in their studies. Further, we discourage the use of excessively high bonuses or punitive penalties. Such practices can pressure participants to prioritize speed or quantity over authenticity and well-being. Researchers should strive to design incentive schemes that offer fair compensation while respecting participant autonomy.

{Secondly, it is important to acknowledge the imbalanced power dynamic between researchers and participants, and how our contributions can be enhanced to bridge the gap. In the future, it is crucial to incorporate participant input and feedback into the design and evaluation of the framework and repository itself. By doing so, we can empower participants to have a larger say in the narrative surrounding their motivations and pay structure.}

Finally, our focus on monetary incentives inherently limits this work to the realm of extrinsic motivation. There is a substantial body of research that acknowledges the importance of intrinsic motivations in driving crowd worker participation as well \cite{IntrinsicMotiv14, IntrinsticExtrinsic18, MonetaryandSocialRewards19}. Future research could investigate the interplay between intrinsic and extrinsic motivations for crowd worker behaviour, potentially revealing a more holistic picture of what drives them. 
As the research field of human-AI collaboration and decision-making evolves, so too will the ethical considerations surrounding incentive design. 
By carefully considering these ethical implications, researchers can utilize the {framework} to design incentive schemes that are not only effective in motivating participants but also uphold ethical principles and contribute to responsible research practices.

%% file: 7_Conclusions.tex
\section{Conclusions}
In conclusion, this work weaves a compelling narrative around the critical role of incentive design in unlocking the potential of human-AI collaboration in decision-making. We began by unraveling the existing research, meticulously examining the landscape of challenges and opportunities through a thematic analysis of existing literature. This understanding provided the foundation for crafting the Incentive-Tuning {Framework} - which aims to empower researchers to design appropriate incentive schemes for their studies. Finally, we provided valuable reporting and documentation tools, driven by the {framework} and its theoretical underpinnings, ensuring accessibility and potential for future refinement. In painting the picture for \textit{understanding}, \textit{designing}, and \textit{documenting} incentive schemes and discussing the implications of our findings and suggestions, we have advocated for a more standardised (yet flexible) approach to the entire incentive design process. This, in turn, can pave the way for more reliable and generalizable knowledge in the field of human-AI decision-making. Ultimately, this journey aims to empower researchers to develop effective human-AI partnerships, leveraging the strengths of both humans and machines, to achieve quality decision-making outcomes across various domains.